\documentclass[letterpaper,journal]{IEEEtran}

\usepackage{amssymb,amsmath,amsthm}
\usepackage[nobreak,noadjust]{cite}

\usepackage{graphicx}
\usepackage{multirow}
\usepackage{bbm}
\usepackage{enumitem}
\setlist[itemize]{leftmargin=*}
\usepackage[english]{babel}

\theoremstyle{definition}
\newtheorem{MAIN-PROB}{Problem}
\newtheorem{CON-B1-RS}{Proposition}
\newtheorem{CON-H2-RS}[CON-B1-RS]{Proposition}
\newtheorem{MAX-RS}{Theorem}
\newtheorem{ETA-BND}{Remark}
\newtheorem{SEC-PROB}[MAIN-PROB]{Problem}
\newtheorem{TRD-PROB}[MAIN-PROB]{Problem}

\begin{document}
\title{Artificial-Noise-Aided Secure Multi-Antenna Transmission with Limited Feedback}
\author{Xi Zhang, \IEEEmembership{Member, IEEE,} Matthew R. McKay, \IEEEmembership{Senior Member, IEEE,}\\
Xiangyun Zhou, \IEEEmembership{Member, IEEE,} and
Robert W. Heath Jr., \IEEEmembership{Fellow, IEEE}

\thanks{X. Zhang was with the Department of Electronic and Computer Engineering, the Hong Kong University of Science and Technology, and is now with the Communication Technology Laboratory, Huawei Technologies Co., Ltd., both in People's Republic of China (e-mail: panda.zhang@huawei.com). M. R. McKay is with the Department of Electronic and Computer Engineering, the Hong Kong University of Science and Technology, People's Republic of China (e-mail: eemckay@ust.hk). X. Zhou is with the Research School of Engineering, the Australian National University, Australia (e-mail: xiangyun.zhou@anu.edu.au). R. W. Heath Jr. is with the Department of Electrical and Computer Engineering, the University of Texas at Austin, United States of America (e-mail: rheath@utexas.edu).}
\thanks{The work of X. Zhang and M. R. McKay was supported by the Hong Kong Research Grants Council under Grant No. 616312. The work of R. W. Heath was supported by the National Science Foundation under Grant No. NSF-CCF-1218338. This paper was presented in part at the IEEE International Conference on Acoustics, Speech, and Signal Processing, Florence, Italy, May~2014 \cite{Zhang2014A}.}
}

\maketitle

\begin{abstract}
  We present an optimized secure multi-antenna transmission approach based on artificial-noise-aided beamforming, with limited feedback from a desired single-antenna receiver. To deal with beamformer quantization errors as well as unknown eavesdropper channel characteristics, our approach is aimed at maximizing throughput under dual performance constraints---a connection outage constraint on the desired communication channel and a secrecy outage constraint to guard against eavesdropping. We propose an adaptive transmission strategy that judiciously selects the wiretap coding parameters, as well as the power allocation between the artificial noise and the information signal. This optimized solution reveals several important differences with respect to solutions designed previously under the assumption of perfect feedback. We also investigate the problem of how to most efficiently utilize the feedback bits. The simulation results indicate that a good design strategy is to use approximately $20\%$ of these bits to quantize the channel gain information, with the remainder to quantize the channel direction, and this allocation is largely insensitive to the secrecy outage constraint imposed. In addition, we find that $8$ feedback bits per transmit antenna is sufficient to achieve approximately~$90\%$ of the throughput attainable with perfect feedback.
\end{abstract}


\section{Introduction}

As a means of providing enhanced security to wireless networks, physical-layer methods have been receiving considerable attention from the research community \cite{Wyner1975,Csiszar1978,Bloch2011,Zhou2013}. In this context, secure techniques for various contemporary architectures have now been considered, including multi-input multi-output systems, relaying systems, cognitive-radio systems, and large-scale networks (see \cite{Oggier2008,Zhang2012,Mukherjee2013,Pinto2012A,Zhou2011B,Zhang2013C} and references therein). A common assumption in work on wireless physical-layer security is that the transmitter has accurate knowledge of the channel to its desired communicating receiver, and in some cases also perfect knowledge of the channel to the eavesdroppers. In general, knowing the eavesdropper channels appears questionable in practice, particularly if such eavesdroppers are ``passive'' and remain quiet. In addition, typically only an approximation for the desired communication channel may be assumed at the transmitter; due, for example, to practical finite-rate constraints on feedback links (which may be insecure) from the desired receiver to the transmitter.

If multiple antennas are available at the transmitter, intelligent signal processing may be used to provide security against unknown passive eavesdroppers. This idea has been exploited in \cite{Goel2008,Gerbracht2010,GhoGho2011,Li2011,Gerbracht2012,Zhang2013A,Li2013}, where artificial noise is generated in a controlled manner to degrade the eavesdropper's signal reception, while, in theory, not causing additional noise at the desired receiver. The solutions in \cite{Goel2008,Gerbracht2010,GhoGho2011,Li2011,Gerbracht2012,Zhang2013A,Li2013} all assume perfect knowledge of the desired receiver's channel at the transmitter (utilized for appropriate beamformer design). Recently, various contributions have attempted to relax this requirement, allowing for certain channel imperfections at the transmitter. In particular, the study in \cite{Zhou2010,Liu2012,Ng2011,Mukherjee2011,Pei2012,Yang2012} focused on the transmission design and analysis when the channel of the desired receiver is assumed to be known at the transmitter but with unbounded Gaussian errors. This
modeling is most appropriate for specifying errors caused by imperfect channel estimation, but it does not model nor address the problem of practical limited-feedback constraints.

Limited feedback channels are important in practice, particularly in widely-used frequency division duplex (FDD) systems, for which the channel knowledge is typically obtained at the receiver through the use of pilot training, and conveyed to the transmitter through the use of digital feedback. Hence, it is important to study secure transmission design under such conditions with limited feedback constraints. Along this line, only a limited amount of work has been done \cite{Rezki2012,Bashar2011,Lin2011,Sun2011B}. To be specific, \cite{Rezki2012} characterized the ergodic secrecy rate performance for single-antenna systems. With multiple transmit antennas, \cite{Bashar2011} analyzed the secrecy outage probability in slow fading channels without using artificial noise. For fast fading channels, considering beamforming with artificial noise, \cite{Lin2011} provided optimal power allocation that maximizes the ergodic secrecy rate in two asymptotic regions, while~\cite{Sun2011B} derived a lower bound to the ergodic secrecy capacity in integral form and studied the optimal power allocation numerically.

In this paper, we consider a scenario where a multi-antenna transmitter communicates with a desired single-antenna receiver in the presence of a passive eavesdropper, assuming that there exists a limited-rate feedback channel from the desired receiver to the transmitter. Different from the single-antenna system in \cite{Rezki2012}, our transmitter is equipped with multiple transmit antennas, thus providing more flexibility in signal processing. Instead of the traditional beamforming approach in \cite{Bashar2011}, artificial-noise-aided beamforming is adopted in this work to provide enhanced security performance. In contrast to the fast fading channels in \cite{Lin2011,Sun2011B}, we consider a slow fading scenario, and seek to design optimized artificial-noise-aided transmission under suitably chosen outage performance measures (as opposed to ergodic rate). Due to the limited feedback constraint, a key challenge faced in the design of artificial-noise-aided beamforming systems is the artificial noise leakage into the desired communication channel, caused by beamformer quantization errors. To deal with this issue, while also accounting for the lack of eavesdropper channel knowledge, we present a novel optimized rate-adaptive transmission approach aimed to maximize throughput under dual performance constraints. The first is a \emph{connection outage} constraint, which specifies a maximal outage level on the desired communication channel; the second is a \emph{secrecy outage} constraint, which governs the level of security against eavesdropping for a ``worst-case'' scenario with zero thermal noise at the eavesdropper. We develop an adaptive transmission strategy that judiciously selects the wiretap coding parameters (i.e., the coding rate and the rate redundancy for achieving secrecy), as well as the optimal power allocation between the artificial noise and the information signal. Our optimized solution reveals several important differences with respect to solutions designed previously under the assumption of perfect feedback \cite{Zhang2013A}, as well as providing practical engineering design insights and guidelines. These are summarized as follows.

\begin{enumerate}
  \item In terms of maximizing the secrecy throughput, the relative amount of available transmit power to allocate to the information signal and the artificial noise depends on the number of feedback bits and the total power available. With additional feedback bits, the transmitter gains confidence in regards to the desired communication channel, and the optimal strategy is to allocate less power to the information signal and give more to the artificial noise. If the total available power is increased, with limited feedback, the optimal power allocation strategy is to give a larger fraction of this power to the information signal and less to the artificial noise. This is in contrast to the optimal power allocation strategy with perfect knowledge of the desired communication channel (i.e., with unlimited feedback), for which the optimal power split between the information signal and artificial noise tends to be fixed as the transmit power grows large.
  \item In our previous work \cite{Zhang2013A}, we showed that with perfect knowledge of the desired communication channel (i.e., unlimited feedback), the secrecy throughput grows unbounded with increasing transmit power. Here in this paper, we point out that with only limited feedback from the desired receiver, even with an arbitrarily large transmit power, the secrecy throughput remains bounded.
  \item Simulation results indicate that for a given total number of feedback bits, a good ``rule of thumb'' is to allocate roughly $80\%$ of the bits for specifying the channel direction information (CDI), and the remainder for specifying the channel gain information (CGI). This allocation strategy gives near-optimal secrecy throughput performance for a wide range of system parameters, and in particular, its optimality is insensitive to the secrecy outage constraint.
  \item In terms of the number of feedback bits required, for practical finite transmission powers, roughly $8$ feedback bits \emph{per antenna} is sufficient to achieve $90\%$ of the secrecy throughput achievable with perfect knowledge of the desired communication channel (i.e., unlimited feedback).
\end{enumerate}

\section{System Model}
To model a slow-varying rich-scattering environment, we consider Rayleigh fading channels that remain constant for each message transmission and change independently from one transmission to the next. The transmitter is equipped with $N\geq2$ antennas, while the desired receiver and the eavesdropper each has only a single antenna. The received signal at the desired receiver can be written as
\begin{align}
y_d&=\mathbf{h}^H\mathbf{x}+n_d\label{eq:channel_bob}
\end{align}
where $\mathbf{h}\sim\mathcal{CN}\left(\mathbf{0},\mathbf{I}_N\right)$ is the desired communication channel,~$\mathbf{x}$ is the transmitted vector, and $n_d\sim\mathcal{CN}\left(0,\sigma_d^2\right)$ is the thermal noise. The received signal at the eavesdropper is
\begin{align}
y_e&=\mathbf{g}^H\mathbf{x}+n_e\label{eq:channel_eve}
\end{align}
where $\mathbf{g}\sim\mathcal{CN}\left(\mathbf{0},\sigma_g^2\mathbf{I}_N\right)$ is the channel to the eavesdropper. Note that here we did not specify a value for $\sigma_g^2$. Also, the thermal noise level at the eavesdropper is typically unknown. To facilitate a robust secure transmission design, we consider a ``worst-case'' scenario with $n_e=0$.

\subsection{Limited Feedback and Quantization}
In this paper, we consider the situation with limited feedback \cite{Santipach2002,Love2004,Santipach2004,Love2003,Mukkavilli2003,Zhou2005,Jindal2006,Mondal2006,Yoo2007,AuYeung2007,Wu2010} from the desired receiver and no feedback from the eavesdropper. For each channel realization, the transmitter first sends a sequence of training symbols. After receiving these, the desired receiver and the eavesdropper both perform channel estimation to obtain knowledge of their own channel, which is assumed to be perfect. While the eavesdropper does not disclose its channel information, the desired receiver decomposes the obtained channel information $\mathbf{h}$ into the CDI~$\mathbf{ h}/{\|\mathbf{h}\|}$ and the CGI $\|\mathbf{h}\|$, which are important for beamforming and rate-adaptation, respectively. This information is conveyed to the transmitter via $B$ feedback bits: $B_1$ bits to quantize the CDI and $B_2=B-B_1$ bits to quantize the CGI. Note that the CGI is real and positive, thus it can be quantized efficiently using a small number of bits \cite{Yoo2007}. Meanwhile, the CDI is a $N$-dimensional unit-norm complex vector. To quantize the CDI, we choose $2^{B_1}$ unit-norm vectors to form a codebook $\mathcal{C}=\left\{\mathbf{c}_1,\ldots,\mathbf{c}_{2^{B_1}}\right\}$, known at both the transmitter and receiver. An index $
\hat{\ell}=\mathop{\arg\max}_{\ell\in\left\{1,\ldots,2^{B_1}\right\}}\left|\mathbf{c}_\ell^H\mathbf{h}\right|$
is computed by the receiver and fed back to the transmitter. Therefore, $\mathbf{c}_{\hat{\ell}}$ is the quantized CDI available at the transmitter.

A common method for generating the quantization codebook $\mathcal{C}$ (though, used mainly for performance analysis purposes) is to employ random vector quantization (RVQ), which independently selects the codebook entries from a uniform distribution on the unit complex hypersphere \cite{Santipach2004,Zhou2005,Jindal2006}. One can typically achieve better performance, however, with properly designed \emph{deterministic} quantization codebooks, and these have been thoroughly investigated \cite{Love2003,Mukkavilli2003}. In \cite{Love2003}, for example, the codebook design was addressed by relating to the problem of Grassmannian line packing, and the resulting design criterion for a good quantization codebook was to minimize the maximum correlation between any pair of quantization vectors. The quantization codebooks used in this paper are generated following this criterion.

For a codebook $\mathcal{C}$, the quantization cell associated with~$\mathbf{c}_\ell\in\mathcal{C}$ is given by $\mathcal{V}_\ell=\left\{\mathbf{z}|\|\mathbf{z}\|=1,\left|\mathbf{z}^H\mathbf{c}_\ell\right|\geq\left|\mathbf{z}^H\mathbf{c}_j\right|,\forall~j\neq\ell\right\}$.
To facilitate our later analysis, as done in \cite{Love2003,Mukkavilli2003,Zhou2005,Jindal2006,Mondal2006,Yoo2007}, we approximate $\mathcal{V}_\ell$ by:
\begin{align}
\tilde{\mathcal{V}}_\ell=\left\{\mathbf{z}|\|\mathbf{z}\|=1,\left|\mathbf{z}^H\mathbf{c}_\ell\right|^2\geq1-2^{-\frac{B_1}{N-1}}\right\}\label{eq:cell_qca}
\end{align}
where the quantity $2^{-B_1/\left(N-1\right)}$ reflects the maximum quantization error in the CDI. This approximation was first introduced in \cite{Love2003,Mukkavilli2003} and then used in \cite{Zhou2005}. The approximated quantization cell can be viewed as a spherical cap on the unit complex hypersphere. As will be seen later, this quantization cell approximation not only allows one to characterize the distribution of quantization error (see Lemma 6 in \cite{Jindal2006}), but also provides an accurate performance indication for any well-designed quantization codebook \cite{Love2003,Mukkavilli2003,Zhou2005,Jindal2006,Mondal2006,Yoo2007}.

\subsection{Artificial-Noise-Aided Beamforming}
Denote the information signal by $u\sim\mathcal{CN}\left(0,\sigma_u^2\right)$. Define a power allocation ratio $\phi$ as the ratio of the information signal power $\sigma_u^2$ to the total transmit power $P$. Thus, $\sigma_u^2=P\phi$. To confuse the eavesdropper, the transmitter performs artificial-noise-aided beamforming \cite{Goel2008} by aligning the information signal along the informed channel direction and injecting artificial noise in orthogonal directions. To be specific, given~$\hat{\ell}$, the transmitted vector $\mathbf{x}$ in (\ref{eq:channel_bob}) admits:
\begin{align}
\mathbf{x}=\mathbf{c}_{\hat{\ell}}u+\mathbf{W}\mathbf{v}\label{eq:an_beamforming_qca}
\end{align}
where $\mathbf{W}$ is a $N\times(N-1)$ complex matrix with $\left[\mathbf{c}_{\hat{\ell}},\mathbf{W}\right]$ being an orthonormal basis, and $\mathbf{v}\sim\mathcal{CN}\left(\mathbf{0},\sigma_v^2\mathbf{I}_{N-1}\right)$ is the artificial noise vector with $\sigma_v^2=P\left(1-\phi\right)/\left(N-1\right)$. Note that this beamforming strategy does not require knowledge of the CGI.

\subsection{Wiretap Coding and Outages}
Before transmission, the data is encoded using Wyner's well-known wiretap coding scheme \cite{Wyner1975}. The codeword rate and the confidential information rate are denoted by $R_b$ and $R_s$, respectively. The codeword rate $R_b$ is the actual transmission rate of the codewords, while the confidential information rate~$R_s$ is the rate of the embedded secret message, to be sent to the desired receiver. The rate redundancy $R_e:=R_b-R_s$ provides secrecy against eavesdropping. More discussion on code construction can be found in \cite{Thangaraj2007}.

Without the eavesdropper's instantaneous channel information, the maximum confidential information rate with perfect secrecy (i.e., the difference between the channel capacities to the desired receiver and the eavesdropper) is unknown and thereby unachievable. Furthermore, with only quantized channel information of the desired receiver, the widely used performance metric -- \textit{outage probability of secrecy capacity}~\cite{Bloch2011} -- is no longer a suitable performance measure, as it does not lead to any directly applicable wiretap coding scheme. In this case, we appeal to a revised secrecy outage formulation proposed in \cite{Zhou2011C,Yuksel2011} to exploit the statistical knowledge of the quantization error and the eavesdropper's channel. That is, we choose the largest possible codeword rate $R_b$ while keeping the required decoding reliability at the desired receiver, and choose the smallest possible rate redundancy $R_e$ while providing the required security performance against eavesdropping, both from a probabilistic sense. By doing so, we maximize the achievable confidential information rate $R_s=R_b-R_e$, which is delivered to the desired receiver while the risks of decoding error and being eavesdropped are both under control. More specifically, if the channel from the transmitter to the desired receiver cannot support $R_b$, the transmitted message cannot be decoded correctly and we consider this a \emph{connection outage} event. If the channel from the transmitter to the eavesdropper can support a data rate larger than $R_e$, perfect secrecy cannot be achieved and a \emph{secrecy outage} event is deemed to occur. In the next section, we first characterize the connection and secrecy outage probabilities.

\section{Outage Performance Analysis}
In this section, we first analyze the connection and secrecy outage performance of the artificial-noise-aided beamforming scheme with limited feedback.

\subsection{Connection Outage Probability}
The connection outage probability is defined as the probability that the capacity of the desired communication channel falls below a preselected codeword rate $R_b$. Denote the instantaneous CDI by $\mathbf{ d}=\mathbf{h}/{\|\mathbf{h}\|}$. Given a feedback index $\hat{\ell}$, the CDI available at the transmitter is $\mathbf{c}_{\hat{\ell}}$. Define
\begin{align}
\cos^2\theta:=\left|\mathbf{d}^H\mathbf{c}_{\hat{\ell}}\right|^2\label{eq:cos_theta}
\end{align}
to reflect how well the obtained CDI aligns with the exact channel direction.

By (\ref{eq:channel_bob}) and (\ref{eq:an_beamforming_qca}), the signal at the desired receiver is
\begin{align}
y_d=\|\mathbf{h}\|\mathbf{d}^H\mathbf{c}_{\hat{\ell}}u+\|\mathbf{h}\|\mathbf{d}^H\mathbf{W}\mathbf{v}+n_d\nonumber
\end{align}
with the signal-to-interference-plus-noise ratio (SINR)
\begin{align}
\mathrm{SINR}_d&=\frac{\|\mathbf{h}\|^2\left|\mathbf{d}^H\mathbf{c}_{\hat{\ell}}\right|^2\sigma_u^2}{\|\mathbf{h}\|^2\|\mathbf{d}^H\mathbf{W}\|^2\sigma_v^2+\sigma_d^2}\nonumber\\
&=\frac{\|\mathbf{h}\|^2\cos^2\theta\sigma_u^2}{\|\mathbf{h}\|^2\left(1-\cos^2\theta\right)\sigma_v^2+\sigma_d^2}\label{eq:snr_bob_cdi}
\end{align}
which follows (\ref{eq:cos_theta}) and the fact that $\left|\mathbf{d}^H\mathbf{ c}_{\hat{\ell}}\right|^2+\|\mathbf{d}^H\mathbf{W}\|^2=1$. The first term in the denominator comes from the artificial noise that leaks into the desired communication channel due to limited feedback and inaccurate beamforming.

To the transmitter, the quantization error in the obtained CDI is unknown, and the instantaneous SINR (i.e., for a given $\mathbf{h}$) at the desired receiver is random. As such, the connection outage probability can be expressed as
\begin{align}
p_\mathrm{co}\left(R_b,\phi,\mathbf{h}\right)&:=\Pr\left(\log_2\left(1+\mathrm{SINR}_d\right)\leq R_b\right).\nonumber
\end{align}
Since the optimal quantization codebook is generally unknown, in this paper, we consider the quantization cell approximation in (\ref{eq:cell_qca}). Based on this approximation, an approximated distribution for the ``quantization error'' $1-\cos^2\theta$, where~$\cos^2\theta$ is defined in (\ref{eq:cos_theta}), was provided in Lemma 6 of \cite{Jindal2006}. Using this result, by (\ref{eq:snr_bob_cdi}), our connection outage probability $p_\mathrm{co}$ can be approximated by
\begin{align}
&~~~~\tilde{p}_\mathrm{co}\left(R_b,\phi,\|\mathbf{h}\|\right)\label{eq:pco_qca}\\
&=\begin{cases}
0&\!\!\!\mathrm{for}~R_b\!\!\leq\!\! R_1\\
1\!-\!2^{B_1}\!\!\left(\!\frac{\|\mathbf{h}\|^2P\phi-\sigma_d^2\left(2^{R_b}\!-\!1\right)}{\|\mathbf{h}\|^2\left(P\phi+\frac{P\left(1-\phi\right)}{N-1}\left(2^{R_b}\!-\!1\right)\right)}\right)^{N\!-\!1}&\!\!\!\mathrm{for}~R_1\!\!<\!\!R_b\!\!\leq \!\! R_2\\
1&\!\!\!\mathrm{for}~R_b\!\!>\!\!R_2
\end{cases}\nonumber
\end{align}
where
\begin{align}
R_1&=\log_2\left(1+\frac{\|\mathbf{h}\|^2P\phi\left(1-2^{-\frac{B_1}{N-1}}\right)}{\|\mathbf{h}\|^2\frac{P\left(1-\phi\right)}{N-1}2^{-\frac{B_1}{N-1}}+\sigma_d^2}\right)\nonumber\\ R_2&=\log_2\left(1+\frac{\|\mathbf{h}\|^2P\phi}{\sigma_d^2}\right).\nonumber
\end{align}

The first boundary value $R_1$ is the capacity of the desired communication channel with maximum quantization error in the CDI. Therefore, $R_1$ is also the maximum data rate achievable without causing connection outages. The second boundary value $R_2$ represents the capacity of the desired communication channel with perfect channel knowledge at the transmitter side. Note that $\tilde{p}_\mathrm{co}$ and $R_1$ are both functions of the number of feedback bits used for the CDI $B_1$. As $B_1\to\infty$, $R_1\to R_2$ and $\tilde{p}_\mathrm{co}$ approaches a step function at $R_b=R_2$. When the number of feedback bits used for the CDI is not too small (e.g., $B_1\geq3N$), (\ref{eq:pco_qca}) provides an accurate approximation for the actual connection outage probability.

\subsection{Secrecy Outage Probability}
The secrecy outage probability $p_\mathrm{so}$ is defined as the probability that the channel capacity to the eavesdropper exceeds a preselected rate redundancy $R_e$. By (\ref{eq:channel_eve}) and (\ref{eq:an_beamforming_qca}), the received signal at the eavesdropper admits
\begin{align}
y_e=\mathbf{g}^H\mathbf{c}_{\hat{\ell}}u+\mathbf{g}^H\mathbf{W}\mathbf{v}\nonumber
\end{align}
with corresponding signal-to-interference ratio (SIR)
\begin{align}
\mathrm{SIR}_e=\frac{\left|\mathbf{g}^H\mathbf{c}_{\hat{\ell}}\right|^2\sigma_u^2}{\|\mathbf{g}^H\mathbf{W}\|^2\sigma_v^2}.\nonumber
\end{align}
Though the eavesdropper's channel is unknown to the transmitter, by \cite[eq. (5)]{Zhang2013A}, the secrecy outage probability can be computed as
\begin{align}
p_\mathrm{so}\left(R_e,\phi\right)&:=\Pr\left(\log_2\left(1+\mathrm{SIR}_e\right)\geq R_e\right)\nonumber\\
&=\left(1+\left(2^{R_e}-1\right){\left({\frac{\phi^{-1}-1}{N-1}}\right)}\right)^{1-N}\label{eq:pso_eve}
\end{align}
which is independent of the informed channel direction $\mathbf{c}_{\hat{\ell}}$, and thus of the desired communication channel $\mathbf{h}$. Here, as a robust design, we ignored the thermal noise at the eavesdropper. Hence, the SIR at the eavesdropper becomes distance-independent, due to the fact that both the signal and the interference come from the same point of transmission. Therefore, the derived expression for the secrecy outage probability is valid for an eavesdropper located at an arbitrary distance. In the literature \cite{Romero2014,Chae2014}, a zero-noise assumption is also used to represent the scenario where the eavesdropper is located arbitrarily close to the transmitter. If the analysis holds true for this case, as one may expect, it is also valid when the eavesdropper is located at a certain (but unknown) distance from the transmitter, due to distance attenuation. In general, our analysis allows the eavesdropper to be located at an arbitrary distance from the transmitter, with the only constraint that the Rayleigh fading assumption still holds.

\section{Secure Transmission Design}
In the last section, we analyzed the connection and secrecy outage probabilities of artificial-noise-aided beamforming scheme with limited feedback. To guarantee the reliability performance of desired communication and the secrecy performance against eavesdropping, we specify a connection and a secrecy outage constraint, respectively. For a given channel realization, the design target is to maximize the confidential information rate under the dual connection and secrecy outage constraints. By averaging the maximum confidential information rate over all channel realizations, we can then investigate the average secrecy throughout performance.

In this section, considering that the CGI is just a positive scalar, which is relatively easy to quantize, we temporarily assume that the transmitter is well-informed about the CGI. In other words, the channel gain is assumed to be accurately known at the transmitter. We then focus on the secure transmission design with quantized CDI and will consider quantization for the CGI in the next section. The main problem we study can be summarized as follows.

\begin{MAIN-PROB}\label{THM-MAIN-PROB}
What are the optimal transmission design parameters that maximize the confidential information rate, under dual connection and secrecy outage constraints?
\end{MAIN-PROB}
For a given realization of the desired communication channel $\mathbf{h}$, recalling that the confidential information rate is given by the difference between the codeword rate and the rate redundancy $R_s=R_b-R_e$, Problem \ref{THM-MAIN-PROB} can be expressed as
\begin{align}
R_s^*\left(\mathbf{h}\right)&=\max_{R_b,R_e,\phi}{\left[R_b-R_e\right]^+}\nonumber\\
&\mathrm{s.t.}~\tilde{p}_\mathrm{co}\left(R_b,\phi,\|\mathbf{h}\|\right)\leq\sigma,~p_\mathrm{so}\left(R_e,\phi\right)\leq\epsilon\label{eq:rs_opt}
\end{align}
where $\left[x\right]^+=\max\left\{0,x\right\}$ and $\sigma,\epsilon\in\left[0,1\right]$ are the enforced connection and secrecy outage constraints. Since an exact expression for $p_\mathrm{co}$ seems unavailable, here we use its approximation $\tilde{p}_\mathrm{co}$, provided in (\ref{eq:pco_qca}). Meanwhile, $p_\mathrm{so}$ was derived in (\ref{eq:pso_eve}) without any approximation. Here we point out that with our design strategy, the \textit{outage probability of secrecy capacity} \cite{Bloch2011}, i.e., $\Pr\left\{C_s<R_s^*\right\}$, where $C_s$ is the unknown secrecy capacity given by the capacity difference between the desired receiver and the eavesdropper, is also guaranteed to be smaller than $\sigma+\epsilon$. This can be proved by invoking the union bound technique and here the proof is omitted for brevity. The solution to the optimization problem in eq. (\ref{eq:rs_opt}) is also the answer to the question we asked in Problem \ref{THM-MAIN-PROB}. That is, solving the optimization problem in eq. (\ref{eq:rs_opt}) will give us the maximum confidential information rate that can be delivered to the desired receiver while the risks of decoding errors and being eavesdropped are both under control.

\subsection{Conditions for Secure Transmission}
The optimization problem in (\ref{eq:rs_opt}) may not always have a positive solution. This occurs when, under dual connection and secrecy outage constraints, the maximum confidential information rate $R_s^*$ is strictly zero. Here, we establish the necessary and sufficient conditions on the system parameters under which a positive $R_s^*$ can be achieved.

Define $\left\lceil x\right\rceil$ as the smallest positive integer which is larger than $x$. We first present a condition on the number of feedback bits used for the CDI.
\begin{CON-B1-RS}\label{THM-CON-B1-RS}
Under the dual outage constraints in (\ref{eq:rs_opt}), for a positive confidential information rate to be achievable, the number of feedback bits used for the CDI must satisfy:
\begin{align}
B_1\geq B_1^{\min}:=\left\lceil\log_2\left(\frac{1-\sigma}{\epsilon}\right)\right\rceil.\label{eq:min_bits}
\end{align}
\end{CON-B1-RS}
\begin{IEEEproof}
See Appendix \ref{PRF-THM-CON-B1-RS}.
\end{IEEEproof}

Here $B_1^{\min}$ is the minimum number of feedback bits, above which a positive confidential information rate is achievable. This minimum requirement is established under the best possible situation where the desired receiver is free of thermal noise, i.e., $n_d=0$. As can be seen from (\ref{eq:snr_bob_cdi}), assuming zero noise for the desired receiver is equivalent to assuming that the desired communication channel has an arbitrarily large strength $\|\mathbf{h}\|$.

From (\ref{eq:min_bits}), we make the following observations:
\begin{itemize}
  \item For a given connection outage constraint $\sigma$, an exponential reduction in the secrecy outage constraint $\epsilon$ necessitates a linear increase in $B_1^{\min}$. Moreover, as $\epsilon\to0$, $B_1^{\min}\to\infty$, implying that as the secrecy outage constraint becomes more and more stringent, an arbitrarily large number of feedback bits is required to achieve any non-zero confidential information rate. This is a consequence of the fact that a more stringent outage constraint $\epsilon$ translates to a larger required rate redundancy $R_e$, and therefore a larger codeword rate $R_b$, for the confidential information rate $R_s=R_b-R_e$ to be positive. With all else fixed, this can obviously be obtained by improving the quality of the desired communication channel through additional feedback bits.
  \item Similarly, for a given secrecy outage constraint $\epsilon$, an exponential reduction in the connection outage constraint $\sigma$ towards zero (i.e., an exponential increase in $1-\sigma$ towards one) necessitates a linear increase in $B_1^{\min}$. Moreover, as $\sigma\to0$, $B_1^{\min}\to\left\lceil\log_2\left(1/\epsilon\right)\right\rceil$, implying that as the connection outage constraint is made more stringent, the required number of feedback bits grows, as above, but in this case it remains bounded. That is, unlike secrecy outages, connection outages can be avoided, provided that the number of feedback bits exceeds this limiting bound. This phenomena, while not immediately obvious, is due to the fact that with a well-designed quantization codebook, the CDI errors due to limited feedback are strictly upper bounded, and these can be made arbitrarily small as the number of feedback bits increases. Indeed, there comes a point in which the codeword rate $R_b$, when matched to the capacity of the desired communication channel under a worst-case quantization noise assumption (thus, avoiding connection outages), can still exceed the required rate redundancy $R_e$, thereby leading to a positive confidential information rate~$R_s$.
  \item Interestingly, the condition in (\ref{eq:min_bits}) shows no dependence at all on the number of transmit antennas $N$. This result is not immediately intuitive, due to the dependence of $N$ on different aspects of the system. To examine this, first recall that this achievability condition is established under the assumption that the desired receiver is free of thermal noise. In this case, as can be seen from (\ref{eq:snr_bob_cdi}), the SINR at the desired receiver depends on the angular mismatch between the beamformer and the channel vector, but not the channel strength. (Thus, the additional array gain available to the beamformer by increasing the number of antennas is inconsequential.) Moreover, increasing $N$ leads to a larger quantization error in the CDI, and this in turn leads to an increased connection outage probability. Consequently, in order to meet the specified connection outage constraint, the transmitter must employ a reduced codeword rate. In contrast to these negative effects, increasing $N$ allows higher dimensionality for the generated artificial noise, which in turn provides additional protection against eavesdropping and thus a reduced secrecy outage probability. In terms of rate, this implies that the transmitter can use a smaller rate redundancy to satisfy the same secrecy outage constraint. What is most curious is that, as implied by (\ref{eq:min_bits}), the reduction of both rates (i.e., $R_b$ due to increased connection outages and $R_e$ due to reduced secrecy outages) as $N$ is increased is the same, such that the difference between them~$R_s$ is unaffected.
\end{itemize}

Table \ref{Table-Feedback-Bits} gives some example values for $B_1^{\min}$ in (\ref{eq:min_bits}) for different connection and secrecy outage constraints. Here, we see that the number of feedback bits required to get a non-zero confidential information rate is generally small.

\setlength{\tabcolsep}{3ex}
\renewcommand{\arraystretch}{1.5}
\begin{table}
  \centering
  \caption{Required Feedback Bits for CDI $B_1^{\min}$}
  \begin{tabular}{c | c | c | c | c | c}
  \hline \hline
  \multicolumn{2}{c |}{\multirow{2}{*}{$B_1^{\min}$}} & \multicolumn{4}{c}{$\epsilon$} \\ \cline{3-6}
  \multicolumn{2}{c |}{} & $1$ & $0.1$ & $0.01$ & $0.001$ \\ \hline
  \multirow{3}{*}{$\sigma$} & $1$ & 1 & 1 & 1 & 1 \\ \cline{2-6}
  & $0.1$ & $1$ & $4$ & $7$ & $10$ \\ \cline{2-6}
  & $0.01$ & $1$ & $4$ & $7$ & $10$ \\ \hline \hline
  \end{tabular}
  \label{Table-Feedback-Bits}
\end{table}

Now we consider the case where the thermal noise at the desired receiver is non-negligible (i.e., $n_d\neq0$). In this case, to achieve a positive confidential information rate, in addition to the number of feedback bits needed to satisfy (\ref{eq:min_bits}), the strength of the desired communication channel must also be sufficiently large, as indicated in the following:

\begin{CON-H2-RS}\label{THM-CON-H2-RS}
Assuming the condition in (\ref{eq:min_bits}) is satisfied, to achieve a positive confidential information rate, the strength of the desired communication channel must also satisfy:
\begin{align}
\|\mathbf{h}\|^2>\mu_{\min}:=\frac{\left(N-1\right)\left(\sqrt[N-1]{\frac{1}{\epsilon}}-1\right)}{P\left(1-\sqrt[N-1]{\frac{1-\sigma}{2^{B_1}\epsilon}}\right)}\sigma_d^2.\label{eq:onoff_threshold}
\end{align}
\end{CON-H2-RS}
\begin{IEEEproof}
See Appendix \ref{PRF-THM-CON-H2-RS}.
\end{IEEEproof}

Here $\mu_{\min}$ is the minimum strength of the desired communication channel required for a positive confidential information rate to be achievable.
From a design perspective, (\ref{eq:onoff_threshold}) implies that one should adopt an ``on-off'' transmission scheme \cite{Zhou2011C}, with a transmit threshold $\mu_{\min}$.

Actually, the condition in (\ref{eq:onoff_threshold}) is closely related to that in~(\ref{eq:min_bits}). Since (\ref{eq:min_bits}) was derived assuming zero thermal noise, it is independent of the strength of the desired communication channel. Now, this fact can also be observed from (\ref{eq:onoff_threshold}). That is, when $n_d=0$ (thus $\sigma_d^2=0$), the requirement on the channel strength disappears (it simply needs to be greater than zero). When the thermal noise at the desired receiver becomes non-negligible (i.e., $\sigma_d^2\neq0$), to compensate for this effect, while the lower bound on the required number of feedback bits does not change, a requirement on the channel strength comes up, as given in (\ref{eq:onoff_threshold}).

One may interpret the interplay between the conditions in~(\ref{eq:min_bits}) and (\ref{eq:onoff_threshold}) from a stochastic point of view. To this end, first note that the CGI $\|\mathbf{h}\|$ is randomly distributed on $\left[0,\infty\right)$. As $B_1$ approaches the theoretical lower limit $\log_2\left[\left({1-\sigma}\right)/{\epsilon}\right]$, the denominator in (\ref{eq:onoff_threshold}) approaches zero and the transmit threshold $\mu_{\min}$ becomes arbitrarily large. Thus, the probability of the event $\{ \|\mathbf{h}\|^2>\mu_{\min}\}$, and consequently the probability of achieving a positive confidential information rate, tends to zero. Clearly, the larger the number of feedback bits $B_1$ (i.e., as it grows beyond its lower limit $\log_2\left[\left({1-\sigma}\right)/{\epsilon}\right]$), the smaller the threshold $\mu_{\min}$, and thus the greater the probability of achieving a positive confidential information rate.

\subsection{Optimized Transmission Design}\label{sec:optimal_design}
Having discussed the conditions under which a positive confidential information rate is achievable, we now provide a closed-form solution to the optimization problem in (\ref{eq:rs_opt}).

\begin{MAX-RS}\label{THM-MAX-RS}
Assume that the conditions in (\ref{eq:min_bits}) and (\ref{eq:onoff_threshold}) are satisfied. The optimal choices of the transmission rates ($R_b$ and $R_e$) and the power allocation ratio $\phi$ in (\ref{eq:rs_opt}) are given by
\begin{align}
\phi^*&\!=\!\frac{\left(\beta\!+\!\sigma_d^2\right)\!\left(\alpha\!-\!\beta\gamma\right)\!-\!\!\sqrt{\alpha\!-\!\beta\gamma\!+\!\sigma_d^2\!\left(1\!-\!\gamma\right)}\!\sqrt{\alpha\gamma\sigma_d^2\!\left(\beta\!+\!\sigma_d^2\right)}}{\beta\!\left(\alpha\!-\!\beta\gamma\right)\!+\!\alpha\sigma_d^2\!\left(1\!-\!\gamma\right)}\nonumber\\
R_b^*&=\log_2\left(1+\frac{\alpha\phi^*}{\beta\left(1-\phi^*\right)+\sigma_d^2}\right)\nonumber\\
R_e^*&=\log_2\left(1+\gamma\frac{\phi^*}{1-\phi^*}\right)\label{eq:phi_opt_qca}
\end{align}
where
\begin{align}
\alpha&=\|\mathbf{h}\|^2P\left(1-\sqrt[N-1]{\frac{1-\sigma}{2^{B_1}}}\right)\nonumber\\
\beta&=\frac{\|\mathbf{h}\|^2P}{N-1}\sqrt[N-1]{\frac{1-\sigma}{2^{B_1}}}\nonumber\\
\gamma&=\left(N-1\right)\left(\sqrt[N-1]{\frac{1}{\epsilon}}-1\right).\label{eq:lux_variables}
\end{align}
The corresponding (strictly positive) maximum confidential information rate is therefore
\begin{align}
R_s^*\left(\|\mathbf{h}\|^2\right)=R_b^*-R_e^*.\label{eq:max_rs}
\end{align}
\end{MAX-RS}
\begin{IEEEproof}
See Appendix \ref{PRF-THM-MAX-RS}.
\end{IEEEproof}

It turns out that for the maximum confidential information rate $R_s^*$ in (\ref{eq:max_rs}), the desired communication channel $\mathbf{h}$ appears only in the form of $\|\mathbf{h}\|^2$. Thus, to facilitate our later analysis, here we have slightly abused notation and written it as~$R_s^*\left(\|\mathbf{h}\|^2\right)$, rather than $R_s^*\left(\mathbf{h}\right)$ as in (\ref{eq:rs_opt}).

\begin{figure}[tb]
  \centering
  \includegraphics[width=0.9\linewidth]{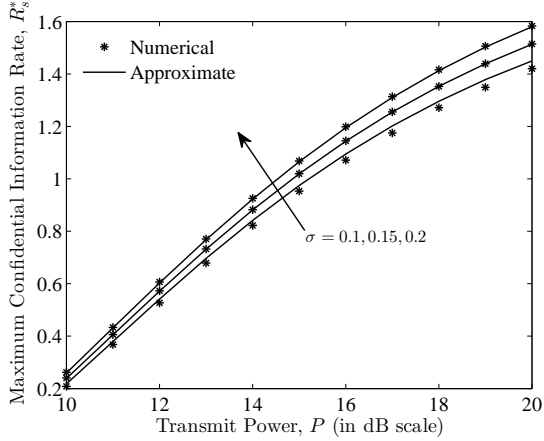}
  \caption{Maximum confidential information rate $R_s^*$ versus the transmit power $P$ for different connection outage constraints $\sigma$. Results are shown for the case where $\sigma_d^2=1$, $N=2$, $B_1=6$, $\epsilon=0.05$ and $\|\mathbf{h}\|=2$.}
  \label{fig:rs_ext_app}
\end{figure}

Note that for the optimization problem in (\ref{eq:rs_opt}), the approximated connection outage probability in (\ref{eq:pco_qca}) was used. Hence, the maximum confidential information rate in (\ref{eq:max_rs}) is also an approximation. In Fig. \ref{fig:rs_ext_app}, we compare (\ref{eq:max_rs}) with the exact maximum confidential information rate, found by inverting the simulated connection outage probability and optimizing the transmit power allocation numerically. The quantization codebooks are generated based on the design criterion in \cite{Love2003,Mukkavilli2003}. As can be seen, the difference between the exact result and our analytical approximation is almost negligible. That is to say, the considered quantization cell approximation has only a negligible effect on the optimality of the proposed transmission design and the solution derived based on the quantization cell approximation accurately predicts the maximum achievable confidential information rate.

\begin{figure}
\centering
\includegraphics[width=0.9\linewidth]{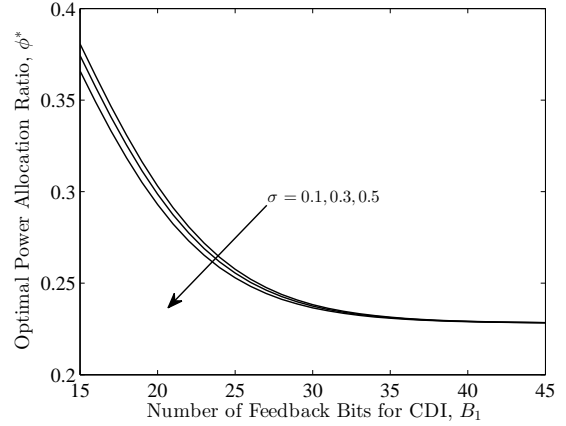}
\caption{Optimal power allocation ratio $\phi^*$ versus the number of feedback bits used for the CDI $B_1$ for different connection outage constraints $\sigma$. Results are shown for the case where $\sigma_d^2=1$, $N=4$, $P=100$, $\epsilon=0.01$ and $\|\mathbf{h}\|=2$.}
\label{fig:phi_opt_b1}
\end{figure}

\begin{figure}
\centering
\includegraphics[width=0.9\linewidth]{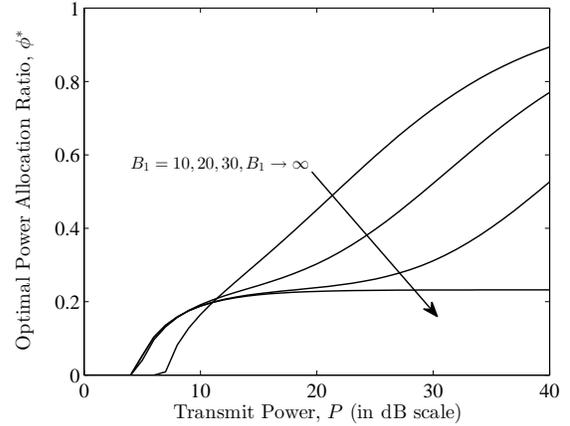}
\caption{Optimal power allocation ratio $\phi^*$ versus the transmit power $P$, compared with the case of accurate CDI. Results are shown for the case where $\sigma_d^2=1$, $N=4$, $\sigma=0.1$, $\epsilon=0.01$ and $\|\mathbf{h}\|=2$.}
\label{fig:phi_opt_p}
\end{figure}

Based on (\ref{eq:phi_opt_qca}), we find the following:
\begin{itemize}
  \item As demonstrated in Fig. \ref{fig:phi_opt_b1}, for a given transmit power $P$, the optimal power allocation ratio $\phi^*$ decreases with increasing number of feedback bits used for the CDI $B_1$. This implies that, if more feedback bits are available, then one should allocate more transmit power to the artificial noise, and less power to the information signal. This may be counter-intuitive, as one could expect the transmitter to give a larger fraction of power to the information signal when it gets more confident about the desired communication channel (the idea of allocating more power to higher quality channels is the basis of conventional waterfilling algorithms, for example). However, with the secrecy constraint, it is the achievable rate \emph{difference} that matters, not only the capacity of the desired communication channel, and the bottleneck affecting the achievable rate difference is the unknown channel of the eavesdropper. With more feedback bits, generating extra artificial noise would have reducing effect on the desired communication channel (approaching asymptotic orthogonality), but could effectively degrade the eavesdropper's signal reception, giving an improved confidential information rate. We also have $\lim_{B_1\to\infty}\phi^*=\phi_\mathrm{perfect}^*$, where $\phi_\mathrm{perfect}^*$ is the optimal power allocation ratio with perfect knowledge of the desired communication channel, derived previously in \cite[eq. (29)]{Zhang2013A}.
  \item As demonstrated in Fig. \ref{fig:phi_opt_p}, for a given number of feedback bits used for the CDI $B_1$, the optimal power allocation ratio~$\phi^*$ increases with the transmit power $P$. This implies that, if extra transmit power is available, then one should allocate more transmit power to the information signal, and less power to the artificial noise. An intuitive explanation can be given as follows. With a fixed number of feedback bits, the desired communication channel has certain advantage over the eavesdropper's channel. By giving a larger fraction of the transmit power to the information signal, this advantage can be further expanded. That is, the increase in the supported codeword rate is larger than that in the required rate redundancy, leading to an improved confidential information rate. Given a finite $B_1$, we also have $\lim_{P\to\infty}\phi^*=1$, implying that with limited feedback and an arbitrarily large transmit power, the confidential information rate is maximized when the transmitter gives all of its power to the information signal. This observation is quite different from the case with perfect knowledge of the desired communication channel, where $\lim_{P\to\infty}\phi_\mathrm{perfect}^*<1$ \cite[eq. (31)]{Zhang2013A}. With limited feedback, as $P\to\infty$, by using a larger power allocation ratio, both the supported codeword rate and the required rate redundancy increase unbounded. However, the difference between them, i.e., the confidential information rate, increases to a finite asymptote. With perfect knowledge of the desired communication channel, as $P\to\infty$, if a certain fraction of the transmit power is reserved for generating artificial noise to limit the eavesdropper's signal reception and thereby the required rate redundancy, the confidential information rate can be made arbitrarily large. This is why different power allocation strategies are observed in the asymptotic region where $P\to\infty$.
\end{itemize}

\subsection{Secrecy Throughput Performance}\label{sec:secrecy_throughput}
In the last subsection, for a given realization of the desired communication channel, we maximized the data rate which is delivered to the desired receiver while the risks of decoding error and being eavesdropped are under control. In this subsection, we investigate the secrecy throughput performance. To be specific, the secrecy throughput is defined as the confidential information rate averaged over all channel realizations, taking into account the probability of connection outage $\sigma$, given by
\begin{align}
\eta=\left(1-\sigma\right)\mathbb{E}_{\mathbf{h}}\left[R_s^*\left(\|\mathbf{h}\|^2\right)\right]\quad\mathrm{(bits/channel~use)}.\nonumber
\end{align}

Assuming that the number of feedback bits used for the CDI $B_1$ is chosen to satisfy (\ref{eq:min_bits}) , and with the optimized transmission design provided in Theorem \ref{THM-MAX-RS}, the corresponding secrecy throughput is
\begin{align}
\eta=\left(1-\sigma\right)\int_{\mu_{\min}}^\infty R_s^*\left(z\right)f_{\|\mathbf{h}\|^2}\left(z\right)\mathrm{d}z\label{eq:throughput_integral}
\end{align}
where the transmit threshold $\mu_{\min}$ is defined in (\ref{eq:onoff_threshold}) and for $z>0$, $f_{\|\mathbf{h}\|^2}\left(z\right)={z^{N-1}\mathrm{e}^{-z}}/{\left(N-1\right)!}$.

In \cite[eq. (35)]{Zhang2013A}, we showed that with perfect knowledge of the desired communication channel, the secrecy throughput grows unbounded (logarithmically) with increasing transmit power. With limited feedback, we present the following remark.

\begin{ETA-BND}
From (\ref{eq:phi_opt_qca})--(\ref{eq:throughput_integral}), as the transmit power grows large, the secrecy throughput converges to a finite asymptote:
\begin{align}
\lim_{P\to\infty}\eta=\left(1-\sigma\right)\log_2\left(\frac{{\sqrt[N-1]{\frac{2^{B_1}}{1-\sigma}}-1}}{\sqrt[N-1]{\frac{1}{\epsilon}}-1}\right).\label{eq:eta_max_p_inf}
\end{align}
\end{ETA-BND}

By adding extra feedback bits (i.e., increasing $B_1$), the throughput limit in (\ref{eq:eta_max_p_inf}) will be increased, as one may expect. An explanation for the observed convergence is given as follows:
\begin{itemize}
  \item With perfect knowledge of the desired communication channel, by increasing the transmit power $P$, the optimal power allocation ratio $\phi_\mathrm{perfect}^*$ in \cite[eq. (29)]{Zhang2013A} increases to an asymptote which is strictly less than one. Therefore, the rate redundancy required to satisfy the secrecy outage constraint is limited to a constant value. Meanwhile, the supported codeword rate grows unbounded with increasing $P$, and this is why the secrecy throughput with perfect knowledge of the desired communication channel increases to infinity, as can be seen from \cite[eq. (35)]{Zhang2013A}.
  \item With quantized CDI feedback from the desired receiver, by increasing the transmit power $P$, the optimal power allocation ratio $\phi^*$ in (\ref{eq:phi_opt_qca}) increases towards one. Consequently, the rate redundancy required to satisfy the secrecy outage constraint increases towards infinity (which is not the case with perfect knowledge of the desired communication channel). Though the supported codeword rate also increases with increasing $P$, the difference between the supported codeword rate and the required rate redundancy converges to a certain value. This is why the secrecy throughput with quantized CDI feedback converges to (\ref{eq:eta_max_p_inf}) instead of growing unbounded with increasing $P$.
\end{itemize}

\begin{figure}[tb]
  \centering
  \includegraphics[width=0.9\linewidth]{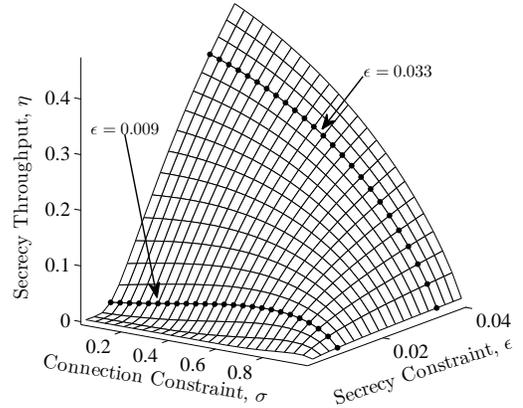}
  \caption{Secrecy throughput $\eta$ versus the connection and secrecy outage constraints $\sigma$ and $\epsilon$. Results are shown for the case where $\sigma_d^2=1$, $N=4$, $P=10$ and $B_1=8$.}
  \label{fig:eta_max_sigma_epsilon_3d}
\end{figure}

Thus far, we have considered the outage constraints, $\epsilon$ and~$\sigma$, as fixed. One may also ask how the specific choice of $\epsilon$ and $\sigma$ influence the throughput performance of our optimized transmission scheme in Theorem \ref{THM-MAX-RS}. This is investigated in Fig.~\ref{fig:eta_max_sigma_epsilon_3d}, which plots the secrecy throughput versus the connection and secrecy outage constraints. As can be seen, for any given connection outage constraint, strengthening the secrecy outage constraint would reduce the achievable throughput, as one may expect. On the other hand, if the secrecy outage constraint is fixed while the connection outage constraint is varied, different behavior is observed depending on the specific value of the secrecy constraint. In particular, if the secrecy constraint is not too strong (e.g., the curve highlighted for $\epsilon = 0.033$), then the secrecy throughput is maximized with as few connection outages as possible; while if the secrecy constraint is sufficiently strong (e.g., the curve highlighted for $\epsilon = 0.009$), then allowing for connection outages (and for this example, quite a lot of outages) can indeed lead to an increased throughput. This observation suggests that a worst-case assumption (i.e., no connection outages allowed, thereby assuming maximum quantization error) does not necessarily yield the maximum secrecy throughput. That is to say, instead of simply assuming maximum quantization error, exploiting the statistical knowledge of the quantization error due to limited feedback can indeed provide a better secrecy throughput.

\section{Quantization of Channel Gain Information}
Up to this point, as in \cite{Jindal2006,Yoo2007}, we have assumed that the CGI is accurately known at the transmitter. In practice, however, this will need to be quantized, along with the CDI. Here we explicitly account for this issue. Specifically, we address the following problem:
\begin{SEC-PROB}
How to quantize the CGI and what is the corresponding secrecy throughput, under dual connection and secrecy outage constraints?
\end{SEC-PROB}

With different numbers of feedback bits allocated to quantize the CGI, we consider different transmission/quantization schemes. Recall that we use $B_2$ feedback bits to quantize the CGI. When $B_2=1$, an on-off transmission scheme is adopted, where the rate parameters and power allocation ratio are chosen to take fixed values, and the feedback bit indicates whether to transmit or not. When $B_2\geq2$, we use a multi-stage quantization scheme to enable adaptive transmission, where the rate parameters and power allocation ratio are chosen based on the feedback about the strength of the desired channel. Each scheme is now discussed in turn.

\subsection{One-Bit CGI Feedback}
When $B_2=1$, this single bit can be used to indicate if the channel strength is above a certain threshold, and thus the transceiver can perform a fixed-rate on-off transmission. More specifically, define $m=\mathbbm{1}_{\left\{\|\mathbf{h}\|^2\geq\mu_T\right\}}$, where $\mathbbm{1}_{\left\{\cdot\right\}}$ is the indicator function and $\mu_T$ is a preselected threshold. The case $m=0$ implies that the transmission should be suspended. Alternatively, if $m=1$, the transmitter conducts transmission using the optimized design provided in Theorem~\ref{THM-MAX-RS}, and assuming that the squared amplitude of the desired communication channel is taking its smallest possible value: $\|\mathbf{h}\|^2=\mu_T$. In this way, the connection and secrecy outage constraints in (\ref{eq:rs_opt}) are both satisfied.

From (\ref{eq:throughput_integral}), the secrecy throughput with one-bit CGI feedback is given by
\begin{align}
\eta=\left(1-\sigma\right)R_s^*\left(\mu_T\right)\tilde{\Gamma}\left(N,\mu_T\right)\label{eq:eta_one_bit}
\end{align}
where $R_s^*\left(\cdot\right)$ was derived previously in (\ref{eq:max_rs}) and $\tilde{\Gamma}\left(\cdot,\cdot\right)$ is the regularized upper incomplete gamma function, defined as $\tilde{\Gamma}\left(N,x\right)=\mathrm{e}^{-x}\sum_{k=0}^{N-1}{x^k}/{k!}$.

As mentioned previously, if $\|\mathbf{h}\|^2\leq\mu_{\min}$, defined in (\ref{eq:onoff_threshold}), then $R_s^*\left(\|\mathbf{h}\|^2\right)=0$. Hence, the transmit threshold should be chosen as $\mu_T>\mu_{\min}$. Note in addition that as $\mu_T\to\infty$, $R_s^*\left(\mu_T\right)$ increases to a finite asymptote, while $\tilde{\Gamma}\left(N,\mu_T\right)$ tends to zero; thus, the throughput in (\ref{eq:eta_one_bit}) also approaches zero. Based on these two observations, we numerically optimize $\mu_T$ in the range $\left(\mu_{\min},\infty\right)$ to maximize the secrecy throughput.

\subsection{Multiple-Bit CGI Feedback}
When $B_2\geq2$, the transceiver can perform rate-adaptive transmission, where the rate choice belongs to a finite set, bounded by the number of quantization steps. In this case, the throughput-optimal CGI quantization scheme is difficult to characterize. Here, we consider a quantization scheme which is reasonably easy to implement and also provides a good throughput performance with a relatively small number of quantization bits.

We first define the inverse function for the regularized upper incomplete gamma function $x=\tilde{\Gamma}_{-1}\left(N,y\right)$ such that $y=\tilde{\Gamma}\left(N,x\right)$. Though there is no closed-form expression for this function, standard software packages such as Matlab provide well-developed routines for numerical evaluation.

From the conditions in (\ref{eq:min_bits}) and (\ref{eq:onoff_threshold}), we know that it is unnecessary to quantize the range $\|\mathbf{h}\|^2<\mu_{\min}$. Meanwhile, though the CGI $\|\mathbf{h}\|$ may take large values, the corresponding probabilities are typically small. Hence, we consider quantization for the CGI in a bounded interval $\mu_1<\|\mathbf{h}\|^2\leq\mu_2$, where $\mu_1=\tilde{\Gamma}_{-1}\left(N,\tilde{\Gamma}\left(N,\mu_{\min}\right)-\delta\right)$, $\mu_2=\tilde{\Gamma}_{-1}\left(N,\delta\right)$, with $\delta$ as a small positive constant. The upper limit $\mu_2$ is introduced to truncate large CGI values appearing with only small probability, while the lower limit $\mu_1>\mu_{\min}$ is invoked to improve the quantization efficiency. (A more detailed explanation of this second condition will be given subsequently.) The value of $\delta$ will be chosen to make sure that $\mu_1<\mu_2$ and to cover an appropriate subset $\left[\mu_1,\mu_2\right]$ of $\left[\mu_{\min},\infty\right)$.

\begin{figure*}[!t]
  \begin{align}
    m\!=\!\begin{cases}
    0\!\!\!&\mathrm{for}~\|\mathbf{h}\|^2\!<\!\mu_1\\
    k\!\!\!&\mathrm{for}~\tilde{\Gamma}_{\!-1}\!\left(\!N,\!\tilde{\Gamma}\!\left(N,\mu_1\right)\!-\!\left(k\!-\!1\right)\!\frac{\tilde{\Gamma}\left(N,\mu_1\right)-\tilde{\Gamma}\left(N,\mu_2\right)}{2^{B_2}-2}\!\right)\!\leq\!\|\mathbf{h}\|^2\!<\!\tilde{\Gamma}_{-1}\!\left(\!N,\!\tilde{\Gamma}\!\left(N,\mu_1\right)\!-\!k\frac{\tilde{\Gamma}\left(N,\mu_1\right)-\tilde{\Gamma}\left(N,\mu_2\right)}{2^{B_2}-2}\!\right)\\
    2^{B_2}\!-\!1\!\!\!&\mathrm{for}~\|\mathbf{h}\|^2\!\geq\!\mu_2\\
    \end{cases}\label{eq:feedback_index}
  \end{align}
  \normalsize \hrulefill
\end{figure*}

The key idea of the considered quantization scheme is to divide the range between $\mu_1$ and $\mu_2$ into $2^{B_2}-2$ quantization intervals, such that $\|\mathbf{h}\|^2$ has the same probability of falling into each interval. This is a quite simple scheme, analogous to the ``histogram equalization'' approach used in image processing \cite{Gonzalez2002}, which we will refer to as ``equalized quantization''. To be precise, for any given CGI $\|\mathbf{h}\|\in\left(0,\infty\right)$, the receiver generates an index $m$ through (\ref{eq:feedback_index}) on the top of the next page, where $k\in\left\{1,\ldots,2^{B_2}-2\right\}$. The value of $m$ is then fed back to the transmitter.

At the transmitter side, when $m=0$ is received, transmission is suspended. When $m\in\left\{1,\ldots,2^{B_2}-1\right\}$ is received, it applies the optimized design provided in Theorem \ref{THM-MAX-RS}, by assuming that the squared amplitude of the desired communication channel is at its smallest value possible, given the quantization value $m$:
\begin{align}
\|\mathbf{h}\|^2\!=\!\tilde{\Gamma}_{\!-\!1}\!\!\left(\!\!N,\tilde{\Gamma}\!\left(N\!,\!\mu_1\right)\!-\!\left(m\!-\!1\right)\!\frac{\tilde{\Gamma}\!\left(N\!,\!\mu_1\right)\!-\!\tilde{\Gamma}\!\left(N\!,\!\mu_2\right)}{2^{B_2}\!-\!2}\!\right)\!\!.\label{eq:assume_gain_equalized}
\end{align}
In this way, the connection and secrecy outage constraints in (\ref{eq:rs_opt}) are both satisfied. This worst-case assumption is also the reason why we choose $\mu_1>\mu_{\min}$: From (\ref{eq:max_rs}), we know that $R_s^*\left(\mu_{\min}\right)=0$. Then, if we were to set $\mu_1=\mu_{\min}$, with the worst-case assumption above, the first quantization interval after $\mu_1$ is wasted. Moreover, choosing $\mu_1$ in excess of $\mu_{\min}$ allows one to impose a lower bound on the achievable confidential information rate $R_s^*\left(\mu_1 \right)$ that must be satisfied, before transmission is conducted. This, in turn, results in more efficient use of the available quantization bits.

With the proposed equalized quantization scheme for the CGI, from (\ref{eq:throughput_integral}), the secrecy throughput becomes
\small
\begin{align}
\eta&=\left(1-\sigma\right)\frac{\tilde{\Gamma}\left(N,\mu_1\right)-\tilde{\Gamma}\left(N,\mu_2\right)}{2^{B_2}-2}\nonumber\\
&\times\!\!\sum_{m=1}^{2^{B_2}\!-\!2}\!\!R_s^*\!\left(\tilde{\Gamma}_{\!-\!1}\!\left(\!N,\tilde{\Gamma}\!\left(N,\mu_1\right)\!-\!\left(m\!-\!1\right)\!\frac{\tilde{\Gamma}\!\left(N,\mu_1\right)\!-\!\tilde{\Gamma}\!\left(N,\mu_2\right)}{2^{B_2}\!-\!2}\right)\!\right)\!\nonumber\\
&+\left(1-\sigma\right)R_s^*\left(\mu_2\right)\tilde{\Gamma}\left(N,\mu_2\right)\nonumber
\end{align}
\normalsize
where $R_s^*\left(\cdot\right)$ is given in (\ref{eq:max_rs}).

\begin{figure}[tb]
  \centering
  \includegraphics[width=0.9\linewidth]{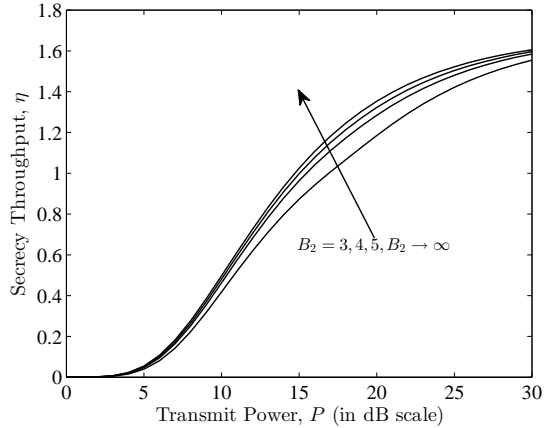}
  \caption{Secrecy throughput $\eta$ versus the transmit power $P$, with quantized CGI, compared with the case of accurate CGI. Results are shown for the case where $\sigma_d^2=1$, $N=4$, $B_1=10$, $\sigma=0.05$, $\epsilon=0.02$ and $\delta=0.0001$.}
  \label{fig:eta_max_p}
\end{figure}

Fig. \ref{fig:eta_max_p} plots the secrecy throughput achieved with the CGI quantization scheme introduced above. Here, the number of bits used for quantizing the CDI $B_1$ is kept fixed, while the number of bits used for quantizing the CGI $B_2$ is varied. We see that with only four or five bits, the equalized quantization scheme provides a throughput performance which is close to that achieved with perfect CGI knowledge (i.e., $B_2\to\infty$).

\subsection{Allocation of Feedback Bits between CDI and CGI}

In practice, a single feedback channel is used for conveying both quantized CDI and CGI. Thus, an interesting problem emerges:
\begin{TRD-PROB}\label{FDBK-DIST-PROB}
For a given number of feedback bits, what is the most efficient allocation between the CDI and CGI in order to maximize the secrecy throughput?
\end{TRD-PROB}

To study this problem, we first define $\tau={B_2}/{B}$, which we term the ``feedback bits allocation ratio''. Clearly, $\tau$ represents the fraction of bits allocated to the CGI, while $1-\tau$ represents the fraction of bits allocated to the CDI. While difficult to characterize Problem \ref{FDBK-DIST-PROB} analytically, we will study this problem through simulations.

\begin{figure}[tb]
  \centering
  \includegraphics[width=0.9\linewidth]{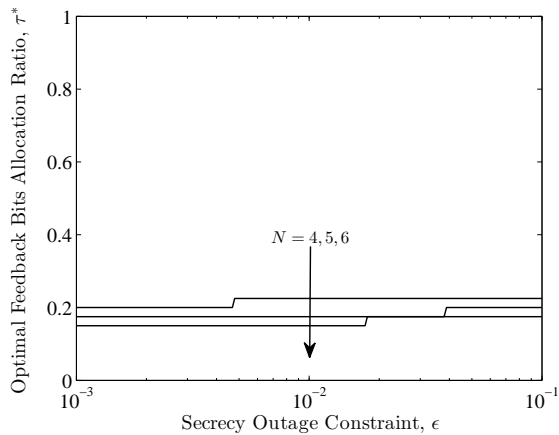}
  \caption{Optimal feedback bits allocation ratio $\tau^*$ versus the secrecy outage constraint $\epsilon$, for different number of transmit antennas $N$. Results are shown for the case where $\sigma_d^2=1$, $P=10$, $B=40$, $\sigma=0.05$ and $\delta=0.0001$.}
  \label{fig:rho_opt_epsilon}
\end{figure}

Intuitively, giving too few feedback bits to either the CGI or the CDI (i.e., with $\tau$ sufficiently close to zero or one respectively) will lead to a degraded secrecy throughput. Fig. \ref{fig:rho_opt_epsilon} shows the optimal $\tau$ that maximizes the secrecy throughput, i.e., $\tau^*$, against the secrecy outage constraint $\epsilon$ for different numbers of antennas $N$. We see that for the scenario considered, $\tau^*\approx0.2$. This number is rather small, but nonetheless intuitive. The CGI is just a positive scalar, which does not require many bits to quantize; while the CDI requires quantization of a $N$-dimensional complex vector (under norm constraints), in order to accurately specify the beamforming direction. Moreover, this optimal allocation ratio is insensitive to changes in the secrecy outage constraint.

Based on the observations above, we claim the following:

\emph{With limited feedback, to maximize the secrecy throughput, a good ``rule of thumb'' is to allocate roughly $20\%$ of the feedback bits for quantizing the CGI, and the remainder for quantizing the CDI.}

In addition to the specific parameterizations considered in Fig. \ref{fig:rho_opt_epsilon}, this claim was also found to be valid over a wider selection of parameters (e.g., $P$, $B$ and $\sigma$). We do not report these additional simulation results here, for conciseness.

\subsection{Quantization Efficiency}
\begin{figure}[tb]
  \centering
  \includegraphics[width=0.9\linewidth]{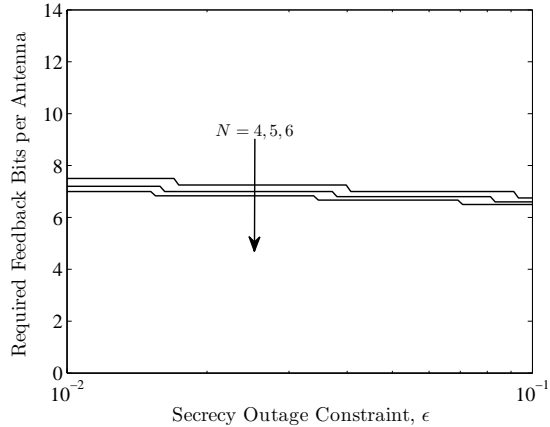}
  \caption{Required number of feedback bits per antenna for achieving $90\%$ of the secrecy throughput with perfect knowledge of the desired communication channel versus the secrecy outage constraint $\epsilon$, for different number of transmit antennas $N$. Results are shown for the case where $\sigma_d^2=1$, $P=20$, $\sigma=0.03$ and $\delta=0.0001$.}
  \label{fig:b_min_90_rho_opt}
\end{figure}

With optimized feedback allocation between the CDI and CGI, we now analyze the quantization efficiency in terms of the number of feedback bits required for achieving a good performance. This issue is studied in Fig. \ref{fig:b_min_90_rho_opt}, which plots the required number of feedback bits to achieve a high fraction (in this example, $90\%$) of the throughput achievable with unlimited feedback. Again, we plot the results as a function of the secrecy outage constraint $\epsilon$, for different antenna numbers~$N$. Based on these results, we make the following interesting observation:

\emph{For a reasonable secrecy outage constraint (e.g., $\epsilon\in\left[0.001,0.1\right]$), roughly $8$ feedback bits per antenna is sufficient for achieving $90\%$ of the secrecy throughput that would be attainable with perfect knowledge of the desired communication channel.}

Once again, although not shown, in addition to the particular parameterizations considered in Fig. \ref{fig:b_min_90_rho_opt}, we also found this claim to be valid for a wider range of parameters (e.g.,~$P$ and~$N$). While Fig. \ref{fig:b_min_90_rho_opt} indicates that $8$ bits of feedback per antenna is sufficient for even reasonably strong outage constraints, as the secrecy outage constraint becomes very relaxed or even removed (e.g., $\epsilon\in\left[0.5,1\right]$), our additional numerical studies have revealed that the required number of feedback bits may indeed be reduced further to around $4$ feedback bits per antenna.

\section{Conclusion and Future Work}
Assuming limited feedback from the desired receiver and no feedback from the eavesdropper, we designed an artificial-noise-aided beamforming technique that enhances secrecy in slow fading channels. Our proposed method was designed to maximize the throughput by adaptively adjusting the wiretap coding rates and the power allocation between the information signal and artificial noise in response to the feedback information, such that dual connection and secrecy outage constraints were met. Our analysis provided key insights into the associated system parameters, such as the optimal power allocation, the number of feedback bits, the number of antennas, the imposed outage constraints, and so on. The proposed method also demonstrated significant differences with respect to previous designs based on perfect feedback.

In this paper, we assumed that the channel estimation at the receiver side is perfect, and focused on the design with a rate-limited feedback link. Possible future extensions include consideration of practical channel estimation schemes with optimized pilot design. In addition, scenarios where the desired receiver and the eavesdropper have multiple antennas are also of interest and will be important topics for future study.

\section{Acknowledgement}
We sincerely thank the anonymous reviewers for their valuable comments which have greatly improved the quality of this paper.

\appendix
\subsection{Proof of Proposition \ref{THM-CON-B1-RS}}\label{PRF-THM-CON-B1-RS}
First note that the connection outage probability $\tilde{p}_\mathrm{co}$ in (\ref{eq:pco_qca}) depends on the codeword rate $R_b$, the secrecy outage probability $p_\mathrm{so}$ in (\ref{eq:pso_eve}) depends on the rate redundancy $R_e$, while both $\tilde{p}_\mathrm{co}$ and $p_\mathrm{so}$ depend on the power allocation ratio $\phi$. For a given $\phi$, the connection outage constraint $\tilde{p}_\mathrm{co}\left(R_b,\phi,\|\mathbf{h}\|\right)\leq\sigma$ implies that the maximum allowable codeword rate is
\begin{align}
R_b^{\max}=\log_2\left(1+\frac{\|\mathbf{h}\|^2P\phi\left(1-\sqrt[N-1]{\frac{1-\sigma}{2^{B_1}}}\right)}{\|\mathbf{h}\|^2\frac{P\left(1-\phi\right)}{N-1}\sqrt[N-1]{\frac{1-\sigma}{2^{B_1}}}+\sigma_d^2}\right).\label{eq:rb_bob}
\end{align}
Similarly, under the secrecy outage constraint $p_\mathrm{so}\left(R_e,\phi\right)\leq\epsilon$, the minimum required rate redundancy is given by
\begin{align}
R_e^{\min}=\log_2\left(1+\frac{\phi}{1-\phi}\left(N-1\right)\left(\sqrt[N-1]{\frac{1}{\epsilon}}-1\right)\right).\label{eq:re_eve}
\end{align}
The confidential information rate $R_s=R_b-R_e$ is therefore maximized when $R_b=R_b^{\max}$ and $R_e=R_e^{\min}$. Clearly, to achieve a positive $R_s$, we require $R_b^{\max}>R_e^{\min}$.

Under the ideal scenario where the desired receiver has zero thermal noise, i.e., $n_d=0$ (thus $\sigma_d^2=0$), $R_b^{\max}$ in (\ref{eq:rb_bob}) admits
\begin{align}
R_b^{\max}|_{n_d=0}\!=\!\log_2\!\left(\!1\!+\!\frac{\phi}{1\!-\!\phi}\!\left(N\!\!-\!1\right)\!{\left(\!\!\!\sqrt[N-1]{\!\frac{2^{B_1}}{1\!-\!\sigma}}\!-\!1\right)}\!\right)\!\label{eq:rb_bob_1}
\end{align}
which is independent of the channel strength $\|\mathbf{h}\|$ and the transmit power $P$. Now, from (\ref{eq:re_eve}) and (\ref{eq:rb_bob_1}), for $R_b^{\max}|_{n_d=0}>R_e^{\min}$ to hold requires ${{2^{B_1}}/{\left(1-\sigma\right)}}>{{1}/{\epsilon}}$,
which is independent of $\phi$ and the number of transmit antennas $N$. Reformulating this condition in terms of the number of feedback bits~$B_1$ and rounding it up to the nearest integer, we establish Proposition \ref{THM-CON-B1-RS}.

\subsection{Proof of Proposition \ref{THM-CON-H2-RS}}\label{PRF-THM-CON-H2-RS}
As discussed in Appendix \ref{PRF-THM-CON-B1-RS}, the confidential information rate $R_s=R_b-R_e$ is maximized when $R_b=R_b^{\max}$ and $R_e=R_e^{\min}$, with $R_b^{\max}$ and $R_e^{\min}$ defined in (\ref{eq:rb_bob}) and (\ref{eq:re_eve}) respectively. These quantities depend on the power allocation ratio $\phi$, which is still to be optimized. In particular, we require
\begin{align}
R_s^*\left(\mathbf{h}\right)&=\max_{0<\phi<1}\left[R_b^{\max}-R_e^{\min}\right]^+.\label{eq:rs_opt_rb_re}
\end{align}

Here, in contrast to Appendix \ref{PRF-THM-CON-B1-RS}, we consider the case where the thermal noise at the desired receiver is non-negligible, i.e., $n_d\sim\mathcal{CN}\left(0,\sigma_d^2\right)$ with $\sigma_d^2>0$. For $R_s^*$ to be positive entails $R_b^{\max}>R_e^{\min}$ for some $\phi$. From (\ref{eq:rb_bob}) and (\ref{eq:re_eve}), an equivalent condition is
\begin{align}
\frac{\|\mathbf{h}\|^2P\left(1-\sqrt[N-1]{\frac{1-\sigma}{2^{B_1}}}\right)}{\|\mathbf{h}\|^2\frac{P\left(1-\phi\right)}{N-1}\sqrt[N-1]{\frac{1-\sigma}{2^{B_1}}}+\sigma_d^2}
>
\frac{N-1}{1-\phi}\left(\sqrt[N-1]{\frac{1}{\epsilon}}-1\right)\nonumber
\end{align}
for some $\phi$, which can be rewritten as
\begin{align}
\|\mathbf{h}\|^2>\frac{\left(N-1\right)\left(\sqrt[N-1]{\frac{1}{\epsilon}}-1\right)}{\left(1-\phi\right)P\left(1-\sqrt[N-1]{\frac{1-\sigma}{2^{B_1}\epsilon}}\right)}\sigma_d^2 \label{eq:min_strength}
\end{align}
for some $\phi$. By letting $\phi=0$ to minimize the right-hand-side, we get the minimum channel strength required for achieving a positive $R_s^*$, reported in Proposition \ref{THM-CON-H2-RS}.

\subsection{Proof of Theorem \ref{THM-MAX-RS}}\label{PRF-THM-MAX-RS}
With the conditions in (\ref{eq:min_bits}) and (\ref{eq:onoff_threshold}) satisfied, by (\ref{eq:min_strength}), the range of the power allocation ratio $\phi$ that gives a positive confidential information rate is
\begin{align}
0<\phi<\phi_{\max}:=1-\frac{\sigma_d^2\left(N-1\right)\left(\sqrt[N-1]{\frac{1}{\epsilon}}-1\right)}{\|\mathbf{h}\|^2P\left(1-\sqrt[N-1]{\frac{1-\sigma}{2^{B_1}\epsilon}}\right)}.\nonumber
\end{align}
Then, the optimization problem in (\ref{eq:rs_opt_rb_re}) reduces to
\begin{align}
R_s^*\left(\mathbf{h}\right)&=\max_{0<\phi<\phi_{\max}}R_b^{\max}-R_e^{\min}\nonumber\\
&=\mathop{\max}_{0<\phi<\phi_{\max}}
\log_2\left(\frac{1+\frac{\|\mathbf{h}\|^2P\phi\left(1-\sqrt[N-1]{\frac{1-\sigma}{2^{B_1}}}\right)}{\|\mathbf{h}\|^2\frac{P\left(1-\phi\right)}{N-1}\sqrt[N-1]{\frac{1-\sigma}{2^{B_1}}}+\sigma_d^2}}
{1+\frac{\phi}{1-\phi}\left(N-1\right)\left(\sqrt[N-1]{\frac{1}{\epsilon}}-1\right)}\right)\nonumber
\end{align}
where $R_b^{\max}$ and $R_e^{\min}$ are defined in (\ref{eq:rb_bob}) and (\ref{eq:re_eve}) respectively. Exploiting the monotonicity of the logarithm function, we solve this by setting the derivative of the quantity inside the brackets to zero, and keeping the only solution which satisfies the constraint. The results obtained are then summarized in Theorem \ref{THM-MAX-RS}.

\bibliographystyle{IEEEtran}
\bibliography{Cited}

\begin{IEEEbiography}[{\includegraphics[width=1in,keepaspectratio]{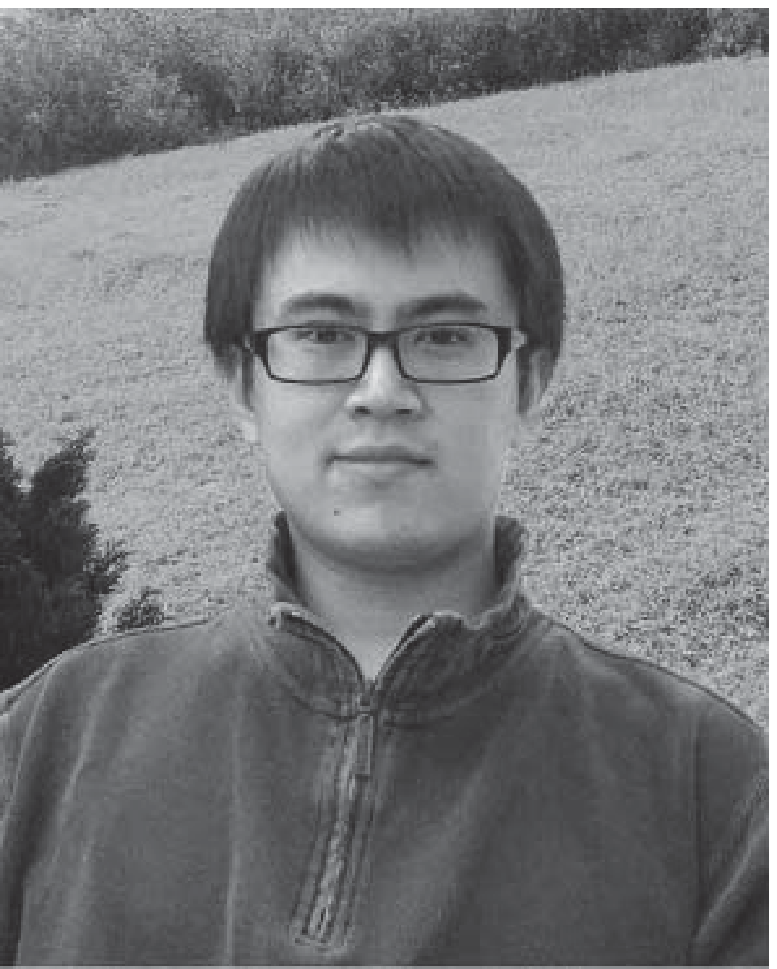}}]{Xi Zhang} (S'11-M'14) received the B.E. degree in communication engineering from the University of Electronic Science and Technology of China in 2010, and the Ph.D. degree in Electronic and Computer Engineering from the Hong Kong University of Science and Technology in 2014. He is now working in the Communication Technology Laboratory, Huawei Technologies Co., Ltd. His current research interests are in the fields of wireless communication with millimeter wave and massive MIMO techniques.
\end{IEEEbiography}

\begin{IEEEbiography}[{\includegraphics[width=1in,keepaspectratio]{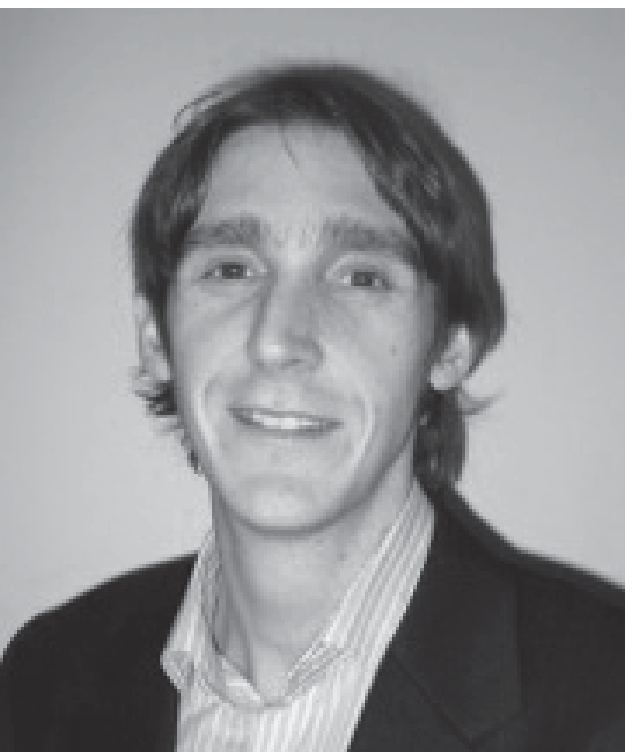}}]{Matthew R. McKay} (S'03-M'07-SM'13) received his Ph.D. from the University of Sydney, Australia, prior to joining the Hong Kong University of Science and Technology (HKUST), where he is currently the Hari Harilela Associate Professor of Electronic and Computer Engineering. His research interests include communications, signal processing, random matrix theory, and associated applications. Most recently, he has developed a keen interest in the interdisciplinary areas of computational immunology and financial engineering.

Matthew and his coauthors have received best paper awards at IEEE ICASSP 2006, IEEE VTC 2006, ACM IWCMC 2010, IEEE Globecom 2010, and IEEE ICC 2011. He also received a 2010 Young Author Best Paper Award by the IEEE Signal Processing Society, the 2011 Stephen O. Rice Prize in the Field of Communication Theory by the IEEE Communication Society, and the 2011 Young Investigator Research Excellence Award by the School of Engineering at HKUST. In 2013, he was the recipient the Asia-Pacific Best Young Researcher Award by the IEEE Communication Society. Matthew is currently serving on the editorial board for the mathematics journal, Random Matrices: Theory and Applications, and he served on the editorial board of the IEEE Transactions on Wireless Communications from 2012 to 2014.
\end{IEEEbiography}

\begin{IEEEbiography}[{\includegraphics[width=1in,keepaspectratio]{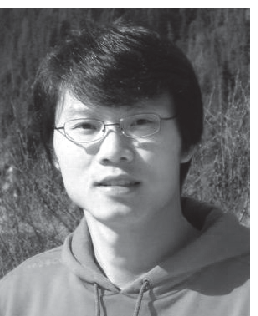}}]{Xiangyun Zhou} (S'08-M'11) received the B.E. (hons.) degree in electronics and telecommunications engineering and the Ph.D. degree in telecommunications engineering from the Australian National University in 2007 and 2010, respectively. From 2010 to 2011, he worked as a postdoctoral fellow at UNIK - University Graduate Center, University of Oslo, Norway. He joined the Australian National University in 2011 and currently works as a Senior Lecturer. His research interests are in fields of communication theory and wireless networks.

Dr. Zhou currently serves on the editorial board of the following journals: IEEE Transactions on Wireless Communications, IEEE Communications Letters, Security and Communication Networks (Wiley). He also served as a guest editor for IEEE Communications Magazine's feature topic on wireless physical layer security and EURASIP Journal on Wireless Communications and Networking's special issue on energy harvesting wireless communications. He was a co-chair of the ICC workshop on wireless physical layer security at ICC'14 and ICC'15. He was the chair of the ACT Chapter of the IEEE Communications Society and Signal Processing Society from 2013 to 2014. He is a recipient of the Best Paper Award at ICC'11.
\end{IEEEbiography}

\begin{IEEEbiography}[{\includegraphics[width=1in,keepaspectratio]{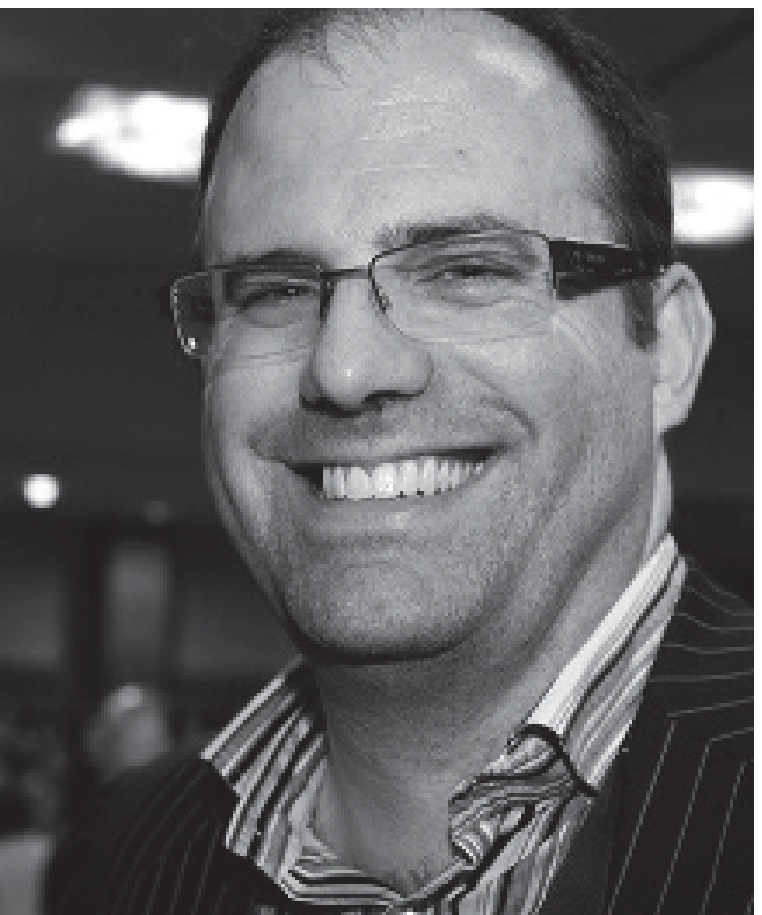}}]{Robert W. Heath Jr.} (S'96-M'01-SM'06-F'11) received the B.S. and M.S. degrees from the University of Virginia, Charlottesville, VA, in 1996 and 1997 respectively, and the Ph.D. from Stanford University, Stanford, CA, in 2002, all in electrical engineering. From 1998 to 2001, he was a Senior Member of the Technical Staff then a Senior Consultant at Iospan Wireless Inc, San Jose, CA where he worked on the design and implementation of the physical and link layers of the first commercial MIMO-OFDM communication system. Since January 2002, he has been with the Department of Electrical and Computer Engineering at The University of Texas at Austin where he is a Cullen Trust for Higher Education Endowed Professor, and is Director of the Wireless Networking and Communications Group. He is also President and CEO of MIMO Wireless Inc. and Chief Innovation Officer at Kuma Signals LLC. His research interests include several aspects of wireless communication and signal processing: limited feedback techniques, multihop networking, multiuser and multicell MIMO, interference alignment, adaptive video transmission, manifold signal processing, and millimeter wave communication techniques. He is a co-author of the book ``Millimeter Wave Wireless Communications'' published by Prentice Hall in 2014.

Dr. Heath has been an Editor for the IEEE Transactions on Communication, an Associate Editor for the IEEE Transactions on Vehicular Technology,  and lead guest editor for an IEEE Journal on Selected Areas in Communications special issue on limited feedback communication, and lead guest editor for an IEEE Journal on Selected Topics in Signal Processing special issue on Heterogenous Networks. He currently serves on the steering committee for the IEEE Transactions on Wireless Communications. He was a member of the Signal Processing for Communications Technical Committee in the IEEE Signal Processing Society and is a former Chair of the IEEE COMSOC Communications Technical Theory Committee. He was a technical co-chair for the 2007 Fall Vehicular Technology Conference, general chair of the 2008 Communication Theory Workshop, general co-chair, technical co-chair and co-organizer of the 2009 IEEE Signal Processing for Wireless Communications Workshop, local co-organizer for the 2009 IEEE CAMSAP Conference, technical co-chair for the 2010 IEEE International Symposium on Information Theory,  the technical chair for the 2011 Asilomar Conference on Signals, Systems, and Computers, general chair for the 2013 Asilomar Conference on Signals, Systems, and Computers, founding general co-chair for the 2013 IEEE GlobalSIP conference, and is technical co-chair for the 2014 IEEE GLOBECOM conference.

Dr. Heath was a co-author of best student paper awards at IEEE  VTC 2006 Spring, WPMC 2006, IEEE GLOBECOM 2006, IEEE VTC 2007 Spring, and IEEE RWS 2009, as well as co-recipient of the Grand Prize in the 2008 WinTech WinCool Demo Contest. He was co-recipient of the 2010 and 2013 EURASIP Journal on Wireless Communications and Networking best paper awards, the 2012 Signal Processing Magazine best paper award, a 2013 Signal Processing Society best paper award, the 2014 EURASIP Journal on Advances in Signal Processing best paper award, and the 2014 Journal of Communications and Networks best paper award. He was a 2003 Frontiers in Education New Faculty Fellow. He is also a licensed Amateur Radio Operator and is a registered Professional Engineer in Texas.
\end{IEEEbiography}

\end{document}